\documentclass[ ]{aastex631}

\pdfoutput=1 
 
\shorttitle{Projection effects in Magnetograms}
\shortauthors{Gosain and Uitenbroek}
\usepackage{graphicx}

\graphicspath{{./}{figures/}}

\begin{document}

\title{Estimation of projection effects in polar magnetic field measurements from the ecliptic view \footnote{Presented as poster in AGU 2021 Fall meeting}}

\author[0000-0002-5504-6773]{Sanjay Gosain}
\affiliation{National Solar Observatory, \\
3665 Discovery Dr.,\\
Boulder, CO 80303, USA}

\author[0000-0002-2554-1351]{Han Uitenbroek}
\affiliation{National SOlar Observatory, \\
3665 Discovery Dr.,\\
Boulder, CO 80303, USA}

\begin{abstract}
The distribution and evolution of the magnetic field at the solar poles through a solar cycle is an important parameter in understanding the solar dynamo. The accurate observations of the polar magnetic flux is very challenging from the ecliptic view, mainly due to (a) geometric foreshortening which limits the spatial resolution, and (b) the oblique view of predominantly vertical magnetic flux elements, which presents rather small line-of-sight component of the magnetic field towards the ecliptic. Due to these effects the polar magnetic flux is poorly measured. Depending upon the measurement technique, longitudinal versus full vector field measurement, where the latter is extremely sensitive to the SNR achieved and azimuth disambiguation problem, the polar magnetic flux measurements could be underestimated or overestimated. To estimate the extent of systematic errors in magnetic flux measurements at the solar poles due to aforementioned projection effects we use MHD simulations of quiet sun network as a reference solar atmosphere. Using the numerical model of the solar atmosphere we simulate the observations from the ecliptic as well as from out-of-ecliptic vantage points, such as from a solar polar orbit at various heliographic latitudes. Using these simulated observations we make an assessment of the systematic errors in our measurements of the magnetic flux due to projection effects and the extent of under- or over estimation. We suggest that such effects could contribute to reported missing open magnetic flux in the heliosphere and that the multi-viewpoint observations from out-of-the-ecliptic plane together with innovative Compact Doppler Magnetographs provide the best bet for the future measurements.
\end{abstract}

\keywords{Sun --- Solar Magnetic Fields --- Activity Cycle --- Interplanetary Magnetic Flux}

\section{Introduction} \label{sec:intro}

\subsection{Motivation:}
 The determination of the distribution and evolution of the polar magnetic flux is important for solar cycle and dynamo studies.
 However, polar observations are challenging from the ecliptic view for the following reasons:
\begin{itemize}
\item Quiet-Sun magnetic flux is predominantly vertical; Oblique view means weak Stokes-V signals.
\item Foreshortening causes coarser spatial sampling per pixel; net flux over polar regions is underestimated due to sub-pixel flux cancellation.
\end{itemize}

Also, the interpretation of the observations itself becomes challenging from oblique view for the following reasons:
\begin{itemize}
\item Spectral lines form over a range of heights in the solar atmosphere.
\item Magnetic field strength decreases with height as flux tube expand upwards to maintain lateral pressure balance.
\item Inclined line-of-sight (LOS) samples different depth in the solar atmosphere compared to vertical LOS.
\end{itemize}

In the present Study we consider the projection effects on the line-of-sight (LOS) measurements. Only the systematic errors are considered; no instrumental noise considered. We use realistic MHD simulations of quiet sun magnetic network as input to synthesize Stokes profiles of the Ni I 676.8 nm absorption line (GONG and MDI heritage).

Stokes profiles are synthesized with various ray projections ($\mu$=0.3-1.0) through the model atmosphere. We then apply center-of-gravity (COG) algorithm to the Stokes I±V profiles to infer “synthetic” LOS magnetic field (BLOS) maps. We then wse the BLOS maps to study net flux variation with viewing angle. Actual Bz at t500=1 surface in the MHD cube is taken as the ground truth.

\subsection{Synthetic Observables:} We used snapshots from the 3D radiative MHD simulations of the solar photosphere produced with the Copenhagen-Stagger code \citep{Galsgaard1996}. Details about these simulations can be found in \cite{Fabbian2015} and references therein. Stokes I and V spectra of the Ni I 676.78 nm spectral line were computed, for the simulated atmosphere under Local Thermodynamic Equilibrium (LTE) approximation, with the Rybicki and Hummer code (RH, \cite{Han2001}), at different inclination angles. In Fig. 1 (top row), we show the synthetic magnetic flux density maps (magnetograms) computed with the center-of-gravity method \citep{Han2003} applied to the Stokes profiles of the Ni I 676.8 nm line. The bottom row shows the computed continuum intensity maps. The viewing angle changes gradually from disk center ($\mu$=1 left-most column) to near limb ($\mu$=0.3 right-most column).

\begin{figure}[ht!]
\begin{center}
\includegraphics[width=150mm]{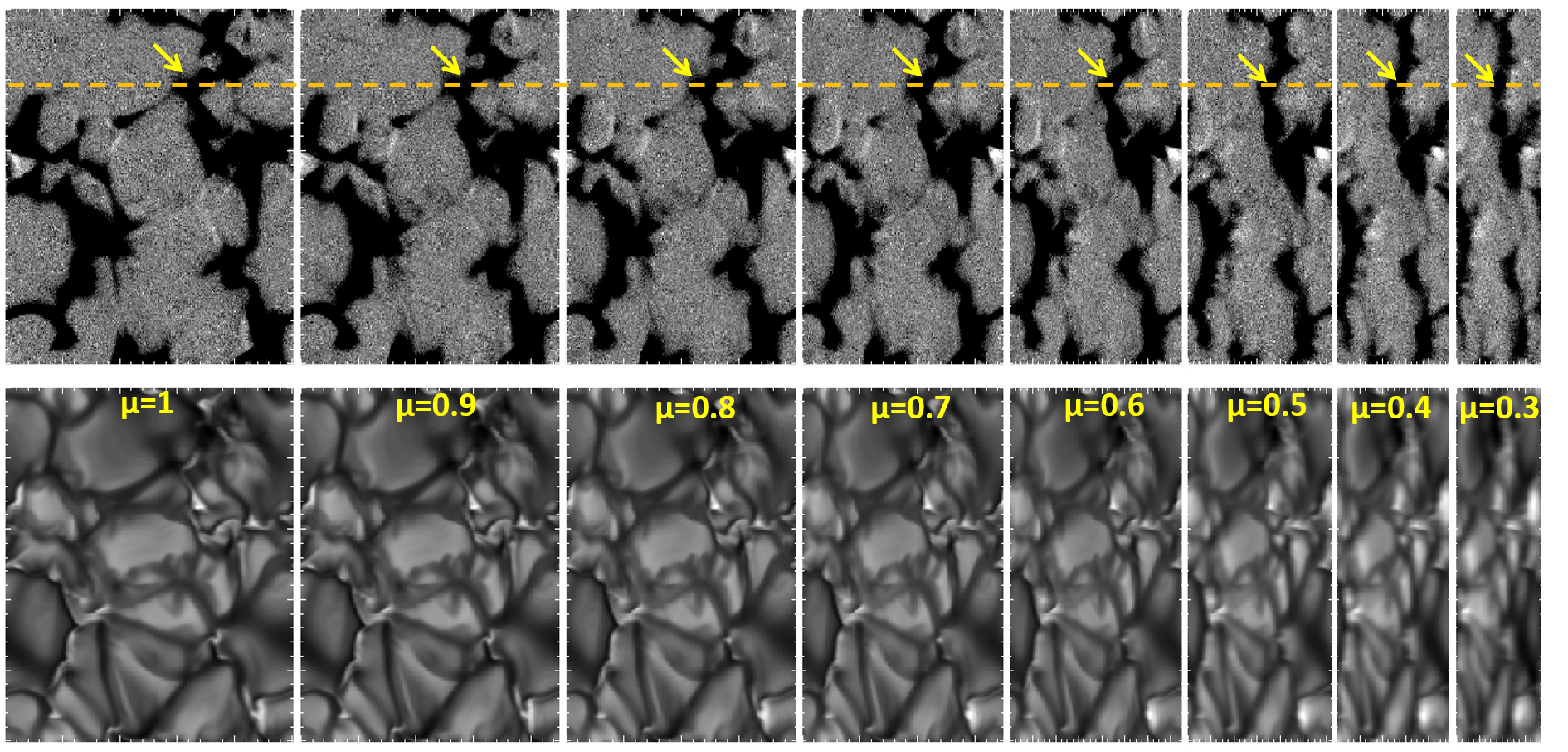}
\caption{A mosaic of synthetic LOS magnetograms (top row) and the continuum intensity map (bottom row) computed for different viewing angles ($\mu$=cos$\theta$), as labeled, from a numerical MHD and radiative transfer model of the quiet Sun network is shown. Geometric foreshortening along x-direction and “bread loaf” like appearance of granules can be noticed (right-most panels in bottom row of the mosaic). The yellow arrow in top row shows the network lane feature which is sampled along horizontal dashed yellow line for further study.  \label{fig:mosaic}}
\end{center}
\end{figure}

\subsection{How does the viewing angle affect the measurements in a magnetic network structure?}
The profile of variation of B-LOS across the horizontal dashed yellow line (so-called "slit") in Figure 1 above is show below. With increasing obliqueness of the LOS we notice:

\begin{itemize}
\item Peak flux density measured in the network feature decreases.
\item The apparent width of the feature increases.
\item The centroid position of the structure shifts. 
\end{itemize}

Further, the absolute value of the net BLOS flux integrated along the "slit" is underestimated by a factor of 5 (solid curve in the inset plot of Fig. 2), when compared $\mu$=0.3 versus $\mu$=1.

Estimating the radial flux along the "slit", BR, under the assumption that the magnetic field is predominantly radial, the BR is underestimated in the feature by a factor of ~1.8, when compared $\mu$=0.3 versus $\mu$=1.

\begin{figure}[ht!]
\begin{center}
\includegraphics[width=150mm]{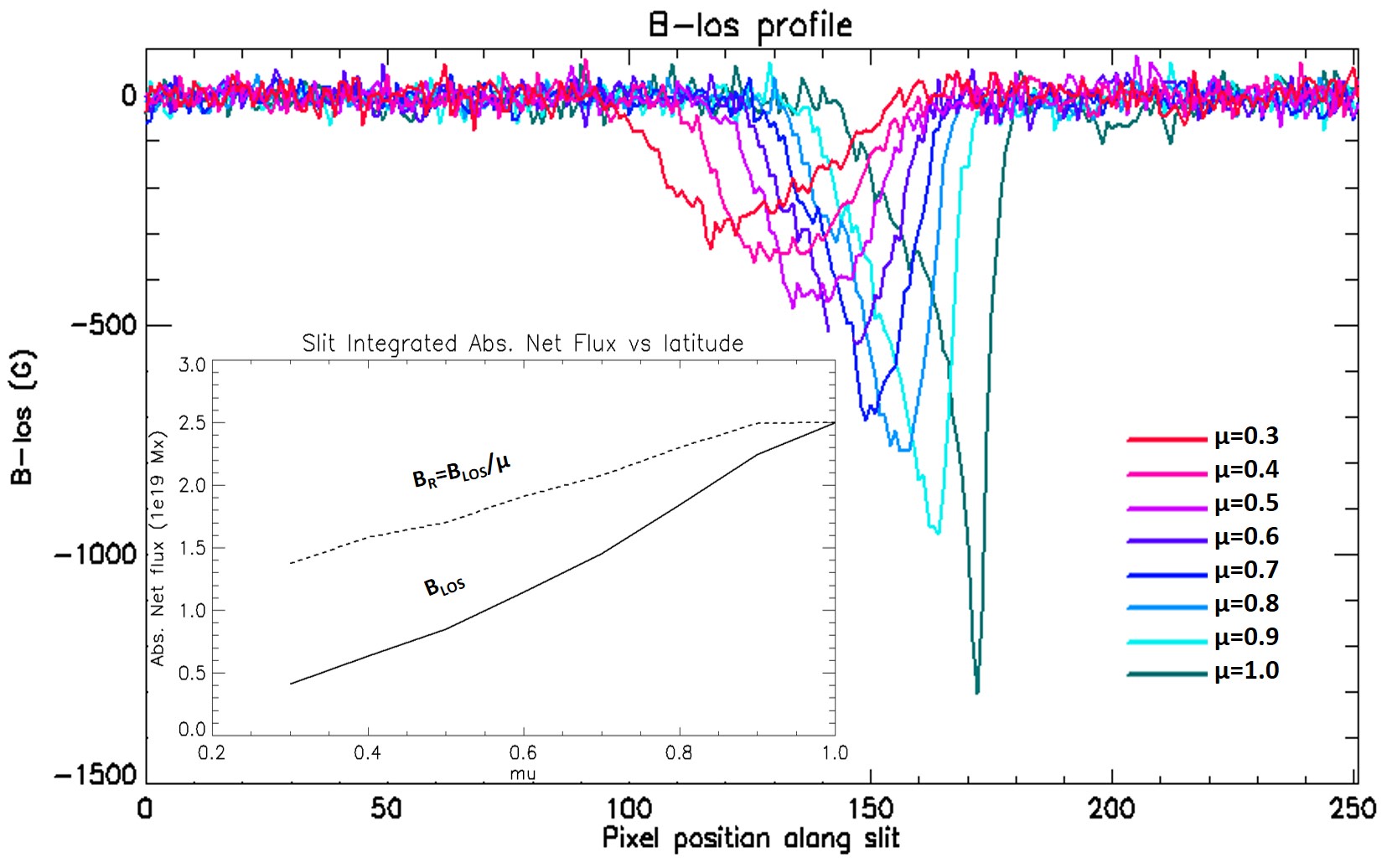}
\caption{The profiles of BLOS variation along the horizontal dashed yellow line (“slit”) in the synthetic magnetograms shown in the mosaic of figure 1, is shown here with the profile colors coded according to the $\mu$ value. The inset shows the variation of the absolute value of net BLOS flux, integrated over the area covered by the slit, with $\mu$ value. Also shown is the variation of the estimated radial flux, BR=BLOS/$\mu$, with $\mu$ value (the dotted curve). \label{fig:slit}}
\end{center}
\end{figure}

\subsection{How does the distribution of Magnetic Flux vary with viewing angle?}

Magnetic flux in the quiet Sun is concentrated in intergranular lanes and is predominantly vertical. It is then expected that with inclined viewing angle the observed LOS component will decrease as cosine of viewing angle or $\mu$. This is seen in the distribution of synthetic magnetograms plotted in top panel of Fig.3.

In principle, if the vertical flux assumption is true everywhere on the Sun then the radial flux can be simply estimated as BR=BLOS/$\mu$, and all the distributions shown in the top panel of Fig 3. should converge.

The bottom panel of Fig. 3 shows the distribution of the radial flux, BR, recovered under the assumption of vertical flux distribution. It can be noticed that:

\begin{itemize}
\item flux distributions of BR for different $\mu$ do not converge, i.e., flux is not vertical everywhere.
\item the stronger flux is not recovered at all from the oblique viewing angles.
\item The weaker positive flux is artificially enhanced in such radial flux estimates.
\end{itemize}

The weaker positive flux presumably comes from the canopy fields expanding towards the solar limb, i.e., away from the LOS.

\begin{figure}[ht!]
\begin{center}
\includegraphics[width=150mm]{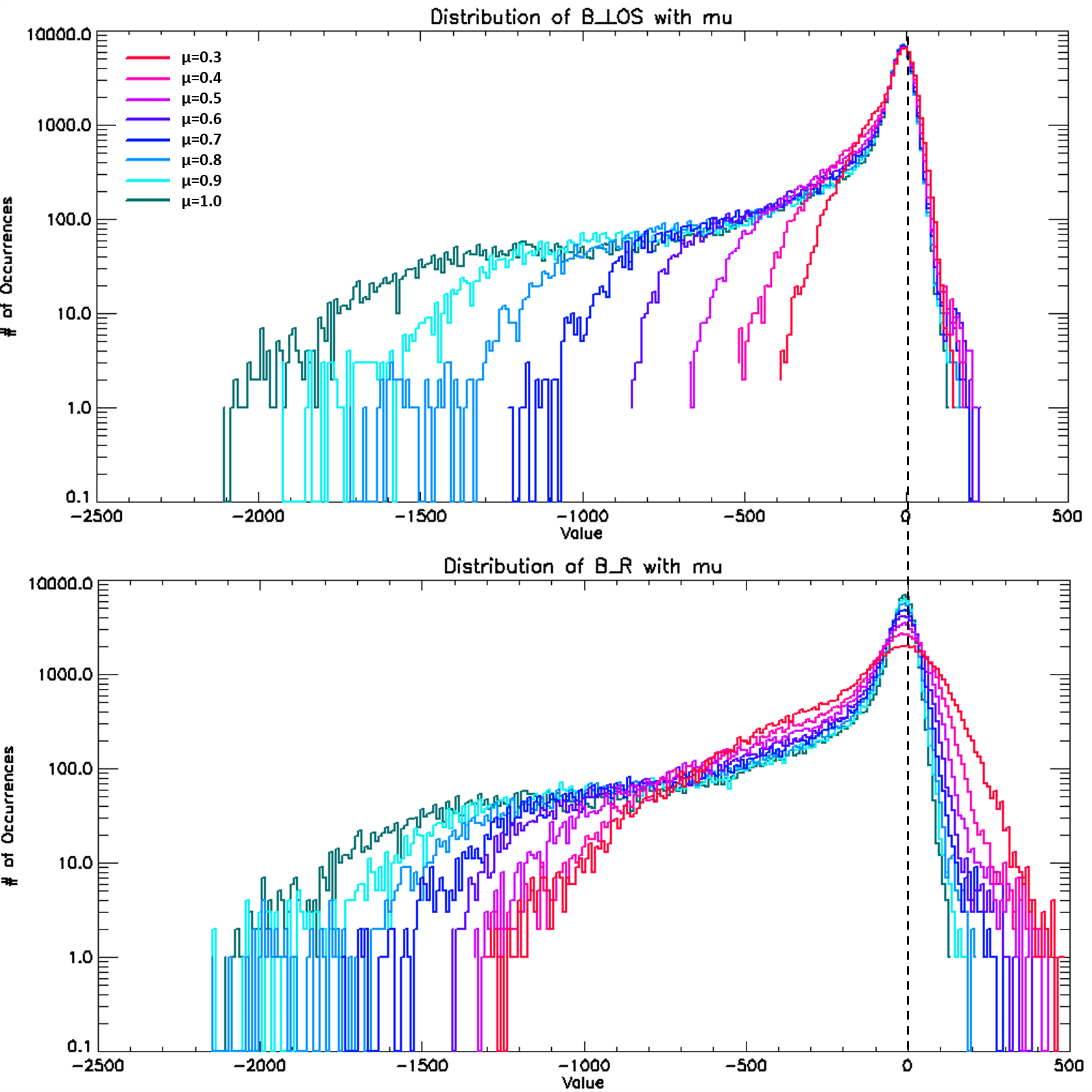}
\caption{ The histograms of the BLOS as inferred from various viewing angles ($\mu$=cos $\theta$) is shown in top panel where $\mu$ is color coded. The number of pixels measuring stronger flux reduces systematically as is expected for the case of predominantly vertical flux tubes in intergranular network region. Traditionally, the radial flux is computed from BLOS as, BR=BLOS/$\mu$, under the assumption that flux is predominantly vertical in the quiet-Sun. The histograms of the BR flux inferred under this assumption at various $\mu$ is shown in the bottom panel. \label{fig:distr}}
\end{center}
\end{figure}

\section{Results}
The mean magnetic flux in solar polar regions inferred from LOS magnetographs is underestimated because of several reasons:\\
(a)The magnetic field is predominantly vertical in quiet Sun network regionsà diminishing Stokes-V signal, thus missing weaker flux below the noise limit of the measurements.\\
(b) Under the assumption that field is predominantly vertical, the radial flux is estimated as BR=BLOS/$\mu$.\\ 

This has the following implications:\\
\begin{itemize}
\item Flux tubes are vertical only in the darkest intergranular lanes, flux tubes expand outside this region to form a magnetic canopy.
\item The canopy fields have a pseudo-bipolar appearance away from disk center, just like unipolar round sunspots appear pseudo-bipolar when viewed away from the center of the disk.
\item Due to foreshortening, a part of this pseudo-bipolar canopy flux would cancel due to sub-pixel cancellation.
\item Even if the canopy flux is spatially resolved, applying a vertical flux assumption to pseudo-bipolar canopy flux causes artificial broadening of histogram FWHM for oblique viewing angles (lower panel of Fig 3.)
\item Performing spatial average (over polar regions) of the pseudo-bipolar canopy flux leads to flux cancellation and hence an underestimation of the polar flux.
\item The changing B-angle of the Sun during ecliptic observations further complicates these measurements.
\end{itemize}

(c) The optical depth unity ($\tau$=1) surface samples deeper parts of the flux tube atmosphere, when viewed near disk center, as compared to inclined LOS which effectively samples progressively higher layers with viewing angle.\\

\section{Conclusions}
With the oblique view of polar regions the magnetic flux is underestimated by the LOS magnetograms due to two compounding effects:
\begin{enumerate}
    \item  LOS samples progressively higher layers of the flux tube atmosphere where the magnetic configuration is no longer dominantly vertical but has more of a canopy-like configuration.
    \item  The deep-rooted stronger vertical flux remains obscured/poorly viewed from the inclined LOS. Unfortunately, the approximation BR=BLOS/$\mu$ is only valid for this vertical part of the flux tube.
    \item  The inclined view samples the canopy region of flux tube giving a pseudo-bipolar appearance which does not contribute to net flux estimates when simply spatially averaged and/or due to foreshortening caused sub-pixel cancellation.
    \item  In principle, vector field measurements should help, however, they suffer from issues such as magnetic fill factor, 180-degree azimuth ambiguity and differential Q, U versus V noise issues.
    \item  Finally, these results indicate that the most accurate polar magnetic flux would be measured from out-of-ecliptic observations, for example, the proposed Solaris mission \citep{Hassler2020} is designed to observe the Sun from the polar vantage points. Even larger mission concepts that target multi-viewpoint observations are the MOST mission \citep{Gopal2024} and the Firefly mission \cite{Nour2023}. 
    \item Such multi-viewpoint missions, which are inherently deep-space missions, will need novel designs for their Doppler magnetograph instruments which can meet the challenge of low mass, low power  and simplicity of measurement (on-board processing) requirements. Such instruments have recently been presented by \cite{Gosain2023} and \cite{Hassler2022}.
    
\end{enumerate}

\bibliography{sample631}{}

\begin{thebibliography}{}
\expandafter\ifx\csname natexlab\endcsname\relax\def\natexlab#1{#1}\fi
\providecommand{\url}[1]{\href{#1}{#1}}
\providecommand{\dodoi}[1]{doi:~\href{http://doi.org/#1}{\nolinkurl{#1}}}
\providecommand{\doeprint}[1]{\href{http://ascl.net/#1}{\nolinkurl{http://ascl.net/#1}}}
\providecommand{\doarXiv}[1]{\href{https://arxiv.org/abs/#1}{\nolinkurl{https://arxiv.org/abs/#1}}}

\bibitem[{{Fabbian} \& {Moreno-Insertis}(2015)}]{Fabbian2015}
{Fabbian}, D., \& {Moreno-Insertis}, F. 2015, \apj, 802, 96,
  \dodoi{10.1088/0004-637X/802/2/96}

\bibitem[{{Galsgaard} \& {Nordlund}(1996)}]{Galsgaard1996}
{Galsgaard}, K., \& {Nordlund}, {\r{A}}. 1996, Astrophysical Letters and
  Communications, 34, 175

\bibitem[{{Gopalswamy} {et~al.}(2024){Gopalswamy}, {Christe}, {Fung}, \&
  {Gong}}]{Gopal2024}
{Gopalswamy}, N., {Christe}, S., {Fung}, S.~F., \& {Gong}, Q. 2024, Journal of
  Atmospheric and Solar-Terrestrial Physics, 254, 106165,
  \dodoi{10.1016/j.jastp.2023.106165}

\bibitem[{{Gosain} {et~al.}(2023){Gosain}, {Harvey}, {Martinez Pillet}, {Hill},
  \& {Woods}}]{Gosain2023}
{Gosain}, S., {Harvey}, J., {Martinez Pillet}, V., {Hill}, F., \& {Woods},
  T.~N. 2023, \pasp, 135, 045001, \dodoi{10.1088/1538-3873/acca49}

\bibitem[{{Hassler} {et~al.}(2020){Hassler}, {Newmark}, {Gibson}, {Duncan},
  {Gosain}, {Harvey}, {Wuelser}, \& {Woods}}]{Hassler2020}
{Hassler}, D., {Newmark}, J.~S., {Gibson}, S.~E., {et~al.} 2020, in AGU Fall
  Meeting Abstracts, Vol. 2020, SH011--0003

\bibitem[{{Hassler} {et~al.}(2022){Hassler}, {Gosain}, {Wuelser}, {Harvey},
  {Woods}, {Alexander}, {Braun}, {Gibson}, {Klein}, {Flores}, {Kohnert},
  {Newmark}, {Osterman}, {Phillips}, {Rose}, {Shoffner}, {Uitenbroek},
  {Heerlein}, {Woch}, \& {Zhao}}]{Hassler2022}
{Hassler}, D.~M., {Gosain}, S., {Wuelser}, J.-P., {et~al.} 2022, in Society of
  Photo-Optical Instrumentation Engineers (SPIE) Conference Series, Vol. 12180,
  Space Telescopes and Instrumentation 2022: Optical, Infrared, and Millimeter
  Wave, ed. L.~E. {Coyle}, S.~{Matsuura}, \& M.~D. {Perrin}, 121800K,
  \dodoi{10.1117/12.2630663}

\bibitem[{{Raouafi} {et~al.}(2023){Raouafi}, {Hoeksema}, {Newmark}, {Gibson},
  {Berger}, {Upton}, \& {Vourlidas}}]{Nour2023}
{Raouafi}, N.~E., {Hoeksema}, J.~T., {Newmark}, J.~S., {et~al.} 2023, in
  Bulletin of the American Astronomical Society, Vol.~55, 333,
  \dodoi{10.3847/25c2cfeb.c647a83d}

\bibitem[{{Uitenbroek}(2001)}]{Han2001}
{Uitenbroek}, H. 2001, \apj, 557, 389, \dodoi{10.1086/321659}

\bibitem[{{Uitenbroek}(2003)}]{Han2003}
---. 2003, \apj, 592, 1225, \dodoi{10.1086/375736}

\end{thebibliography}
\bibliographystyle{aasjournal}

\end{document}